\documentclass[aip, apl, amsmath, amssymb, reprint]{revtex4-1}
\usepackage{amsmath,amssymb,graphicx}
\begin{document}

\title{Tapered nanofiber trapping of high-refractive-index nanoparticles}

\author{Jon D. Swaim}\email{Corresponding author: jswaim@physics.uq.edu.au}
\affiliation{School of Mathematics and Physics, University of Queensland, St Lucia, QLD 4072, Australia}
\author{Joachim Knittel}
\affiliation{School of Mathematics and Physics, University of Queensland, St Lucia, QLD 4072, Australia}
\author{Warwick P. Bowen}
\affiliation{Centre for Engineered Quantum Systems, School of Mathematics and Physics,
University of Queensland, St Lucia, Brisbane, QLD 4072, Australia}

\begin{abstract}
A nanofiber-based optical tweezer is demonstrated.  Trapping is achieved by combining attractive near-field optical gradient forces with repulsive electrostatic forces.  Silica-coated Fe$_2$O$_3$ nanospheres of 300 diameter are trapped as close as 50 nm away from the surface with 810 $\mu$W of optical power, with a maximum trap stiffness of 2.7 pN $\mu$m$^{-1}$.  Electrostatic trapping forces up to 0.5 pN are achieved, a factor of 50 larger than those achievable for the same optical power in conventional optical tweezers.  Efficient collection of the optical field directly into  the nanofiber enables ultra-sensitive tracking of nanoparticle motion and extraction of its characteristic Brownian motion spectrum, with a minimum position sensitivity of 3.4 $\AA / \sqrt{\text{Hz}}$.%  This sensitivity is xx $\times$ lower than in an optical tweezer on account of the nanofiber's strong field confinement, and could provide a simple architecture for sensing applications such as single molecule force spectroscopy.  
\end{abstract}

%\ocis{(000.00000) ocis info} 
%ss
\maketitle %% required

%\section{Introduction}		
% Single-sided fiber probes taken out as they trap microparticles  ~\cite{Liu-2006, Xin-2012} 	 
% Threw out ~\cite{Taylor-2013}. reference for need for low optical power

The sensing and manipulation of small dielectric particles has useful applications in single molecule biophysics~\cite{Greenleaf-2006, Neuman-2008}, label-free biomolecule detection~\cite{Vollmer-2002, Arnold-2009, Armani-2007} and nanotechnology~\cite{Erickson-2009}.  Optical tweezers, one of the most well-known manipulation techniques, can stably trap micro-scale particles with forces compatible with the molecular machinery of living cells~\cite{Jannasch-2012}.  This has enabled force spectroscopy on individual biomolecules such as DNA~\cite{Wang-1997}, RNA~\cite{Bustamante} and motor proteins~\cite{Block}.  Currently, there is much interest in extending the achievable trapping forces into the nanonewton range and broadening the applications of optical tweezers to include processes involving larger biological structures such as protein folding~\cite{Jannasch-2012}.  Although a variety of methods have been developed in order to increase trapping forces~\cite{Jannasch-2012, Simpson-1997, Reihani-2007, Mahamdeh-2011, Bormuth-2008}, most still rely on mW levels of optical power and micro-scale particles.  To minimize intrusion on the biological system, it is important to extend these advances to smaller nano-scale probes and $\mu$W powers.  Unfortunately, the diffraction limit of light and the precipitous scaling of trap strength with probe size limit the ability of optical tweezers to meet these challenges~\cite{Erickson-2009}. 

\begin{figure*}[ht!]
%	\begin{center}
\centering
\includegraphics*[scale=0.8]{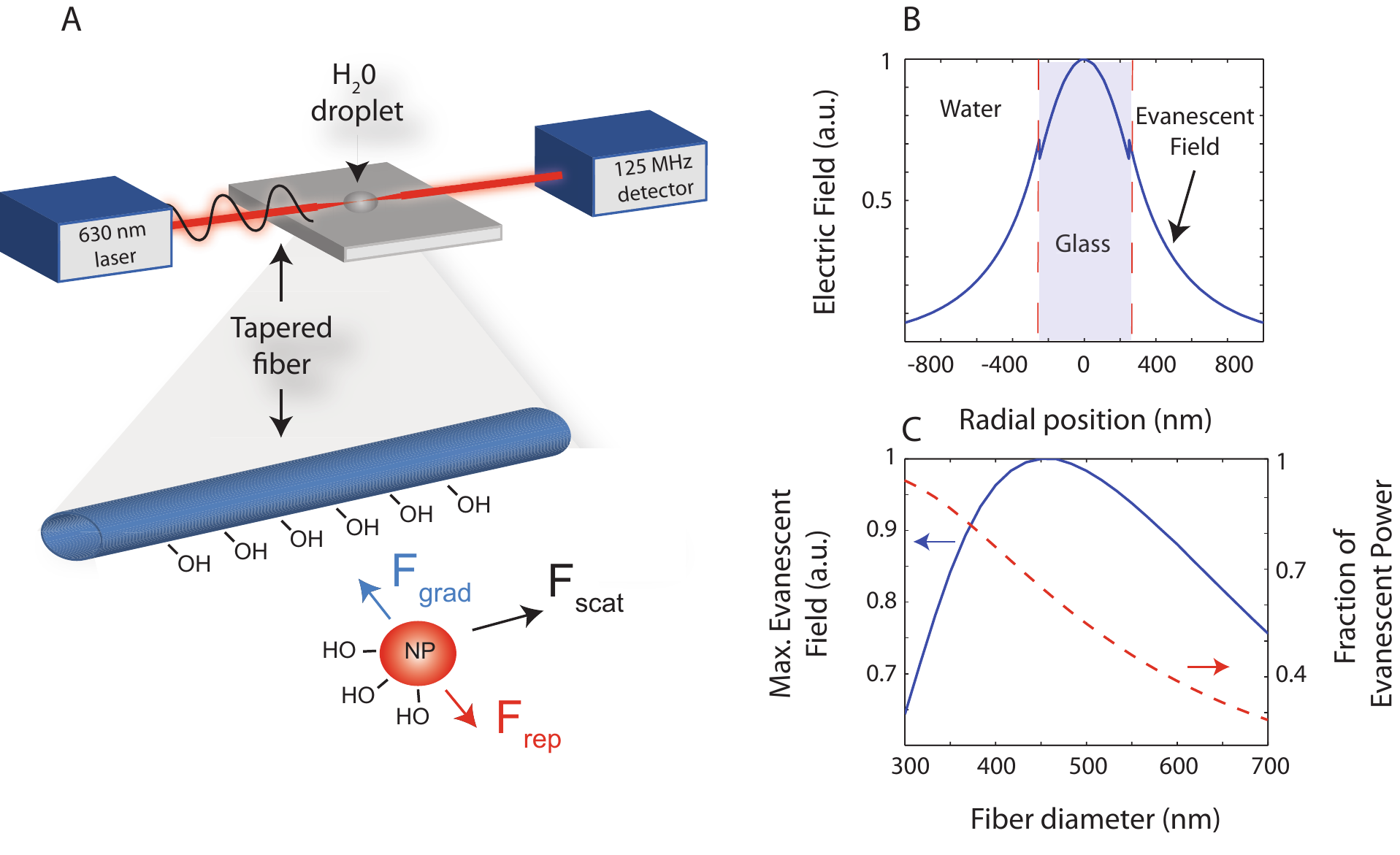}
	\caption{(a) Experimental setup of the nanofiber tweezer.  The surface functionalization of the fiber and nanoparticle (NP) create a repulsive electrostatic force which is opposed by the attractive optical gradient force.  (b) Finite element simulation of radial variation in electric field amplitude for a 500 nm nanofiber.  (c) Electric field amplitude (left) and fraction of evanescent power (right) as a function of various nanofiber diameters.}
	\label{fig:setup}
%	\end{center}
\end{figure*}  

Sub-diffraction limited trapping has been achieved in systems such as slot waveguides~\cite{Erickson-2009}, plasmonic nano-tweezers~\cite{Juan-2011} and optical microresonators~\cite{Arnold-2009, Lin-2010, Mandal-2010}.  In such systems, near-field effects led to stronger forces without the need for large optical power. Here, we extend these approaches to nanofiber-based optical tweezers, demonstrating trapping and tracking of 300 nm silica-coated, high-refractive-index ($n = 2.42$) particles.  Following pioneering work in optical microcavity-based trapping reported in Ref.~\cite{Arnold-2009}, a combination of strong near-field attractive forces and electrostatic repulsive forces traps nanoparticles as close as 50 nm away from the nanofiber surface with only 810 $\mu$W of optical power.  Trap stiffnesses as high as 2.7 pN $\mu$m$^{-1}$ are achieved, as well as peak electrostatic and optical trapping forces of 0.5 pN and 0.14 pN, respectively, which are both more than an order of magnitude larger than those achievable in conventional optical tweezers under similar conditions.  Although tweezers have previously been reported using optical gradient forces from single-sided tapered fibers~\cite{Liu-2006, Xin-2012}, our experiments differ in that they utilize strong electrostatic repulsion and a continuous tapered fiber design with high numerical aperture which enables efficient collection of the scattered light.  This allows ultra-sensitive tracking of nanoparticle motion with a position sensitivity of 3.4 $\AA / \sqrt{\text{Hz}}$. By extending optical tweezers into the nanoscale regime, providing ultra-sensitive tracking and strong forces, as well as eliminating the need for high NA objectives, the technique provides a simple architecture for applications in single molecule force spectroscopy and characterization of molecular interactions, and paves the way towards all-fiber-based optical trapping of nano-scale particles.

As illustrated in Fig.~\ref{fig:setup}(a), the basic principle behind the trapping method implemented here is to balance attractive near-field optical gradient forces with repulsive electrostatic forces~\cite{Arnold-2009}.  In the evanescent field of the nanofiber, the optical gradient force attracts particles towards the position of maximum field intensity at the surface of the fiber, scaling linearly with the injected optical power.  Consequently, using only optical forces, stable trapping is not possible in the field of a nanofiber.  The gradient force can be opposed, however, by electrostatic repulsion between charged surface groups on the particle and the fiber.  The combination of gradient attraction and electrostatic repulsion then produces a potential well in the radial direction where the particle can be trapped.  In the trap, position-dependent scattering from the particle decreases the transmitted power collected at the photodetector~\cite{Zhu-2011}, allowing precise measurement of particle's position provided that the evanescent field profile is known.  Perturbations in the particle's position, such as those resulting from forces applied by nearby biomolecules, could then be easily detected to perform force spectroscopy on individual biomolecules present in between the nanofiber surface and the trapped particle. 

The scattered power from a single nanoparticle is given by the product of the incident field intensity with the particle's scattering cross-section $\sigma_s$:
\begin{equation}
P_{\text{scat}}(r_p) = \frac{1}{2} c \epsilon_0 |E_0(r_p)|^2 \sigma_{\text{s}}
\label{eq:rad_power}
\end{equation}
where $c$ is the speed of light, $\epsilon_0$ is permittivity of free space and $|E_0(r_p)|^2$ is the squared modulus of the electric field at the particle position $r_p$.  To maximize the electric field magnitude at the surface of the nanofiber, the electric field distribution of the HE$_{11}$ mode of a nanofiber in water was calculated as a function of diameter using finite element modeling software COMSOL Multiphysics 3.4, with the result for a 500 nm diameter fiber shown in Fig.~\ref{fig:setup}(b).  As the diameter of the nanofiber is decreased, the fraction of evanescent power increases (see red dashed curve in Fig.~\ref{fig:setup}(c)).  However, the width of the optical mode also increases and consequently the maximum evanescent amplitude occurs at an intermediate nanofiber diameter.  As shown in Fig.~\ref{fig:setup}(c), the model predicts the evanescent amplitude to be within 5\% of the maximum over a range of diameters from 390 to 570 nm, with the maximum occurring at approximately 460 nm.  In this work,  nanofibers 500 nm in diameter were fabricated by stretching standard 630 nm single-mode fiber under a H$_2$ torch until the measured transmission was single-mode and the diameter reached 500 nm as confirmed by scanning electron microscopy.  

Fig.~\ref{fig:setup}(a) illustrates the experimental trapping setup.  Light from a fiber-coupled laser source ($\lambda$ = 630 nm) was guided into the tapered nanofiber and the transmitted light was collected on a photodetector and analyzed on an oscilloscope at a sampling rate of 5 kHz.  By bringing the nanofiber into close proximity with a glass coverslip, a $\mu$L-sized droplet of pure water completely immerses the nanofiber as shown in Fig.~\ref{fig:setup}(a).  A dilute solution of 300 nm silica-coated Fe$_2$O$_3$ nanospheres (density of 200 $\mu$g/cm$^3$;  n$_p$ = $\sqrt{\epsilon_p}$ = 2.42) was then added to the droplet, and the nanofiber’s transmitted light was recorded on the oscilloscope.

\begin{figure*}[ht!]
	\begin{center}

	\includegraphics*[width=1.8\columnwidth]{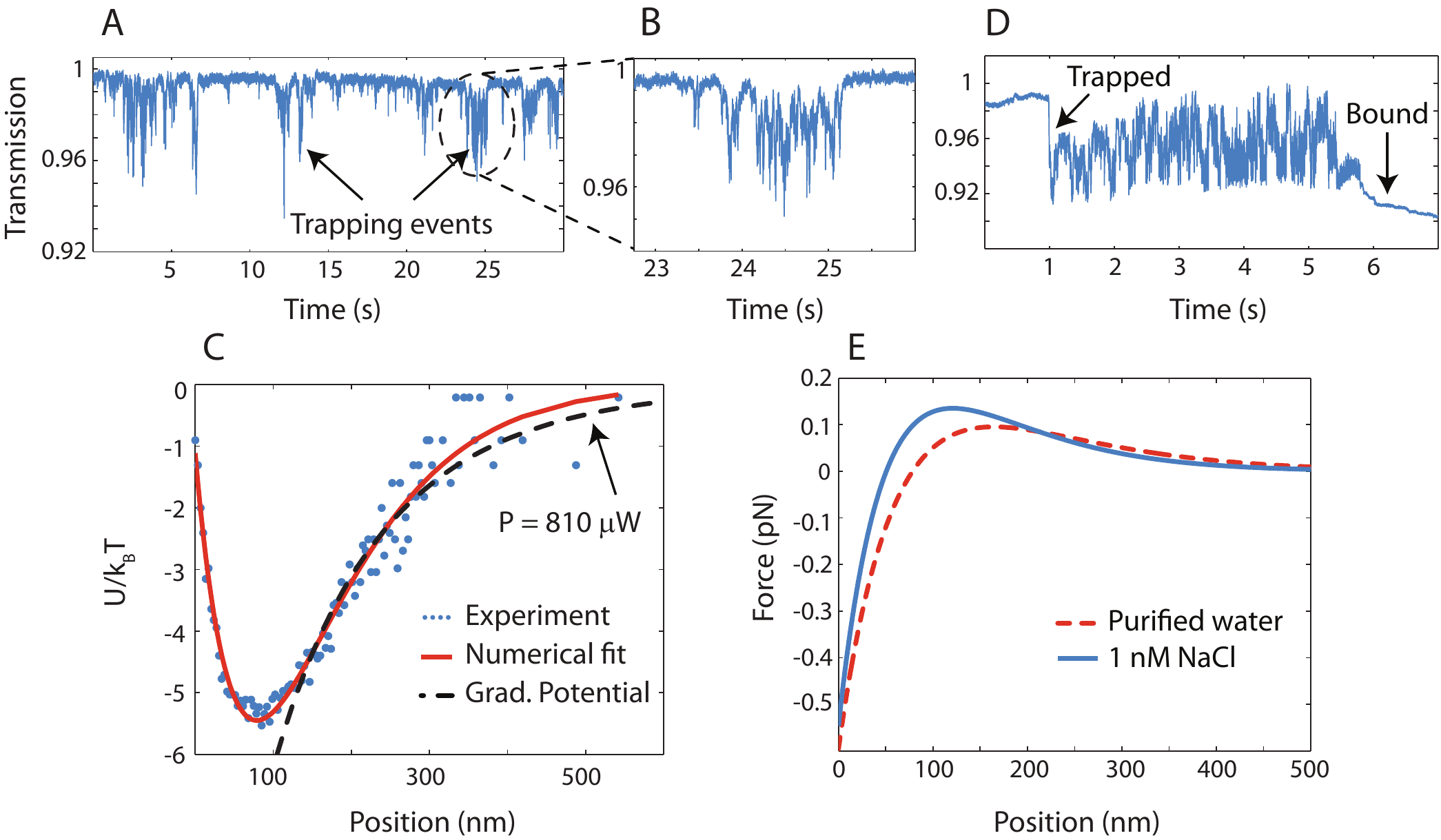}
	\caption{(a) Normalized nanofiber transmission in purified water.  Discrete drops in transmission correspond to single nanoparticle trapping events.  (b) Magnified image of a trapping event at 24 seconds.  (c) Measured (blue dots) potential $U(r_p)/k_B T$, numerical fit (solid red trace) and expected gradient potential (dashed black curve) based on finite element modeling.  (d) Trapping event in 1 nM NaCl solution.  (e) Trapping forces on a single nanoparticle in purified water and 1 nM NaCl solution.}
	\label{fig:nanospheres}
	\end{center}
\end{figure*}

  %Assuming that the nanoparticle interacts with the fiber at the position of maximum intensity (where $r=a/2$), we estimate that the maximum scattered power of a single nanoparticle ($n_p = \sqrt{\epsilon_p} = 2.42$) should be about x nW.

%In this work we use silanol-functionalized particles, which should have a comparable density of silanol groups to that of the bare nanofiber ($\sim$ 10$^{14}$ cm$^{-2}$)~\cite{Vollmer, other}.   %During this time, the nanoparticle is trapped by the opposition of the repulsive electrostatic force and the attractive gradient force $F_{\text{grad}}(r_p)=-1/4 \epsilon_b \alpha \nabla |E(r_p)|^2$.  

%$U_{\text{grad}}(r_p)= -1/4 \epsilon_b \alpha |E(r_p)|^2$.   $U_{\text{grad}}(r_p)= \int_{\infty}^{r_p} F_{\text{grad}} \text{dr}$ 

Fig.~\ref{fig:nanospheres}(a) shows the detected light transmitted through the nanofiber after addition of the nanoparticles.  Sharp drops in the tranmission are seen as the nanoparticles enter the evanescent field of the fiber and are trapped by the combined electrostatic and optical forces.  Ultimately, however, the nanoparticles are repelled from the fiber surface due to electrostatic repulsion.  The total duration of this experiment was 155 seconds, in which about 130 trapping events occured before evaporation of the liquid droplet was evident in the nanofiber's transmitted light.  Fig.~\ref{fig:nanospheres}(b) shows a magnified image of a single trapping event.  The scattered power fluctuates over a duration of about 1 second before the nanoparticle diffuses away.  By considering the position dependence of the scattered power (Eq.~\ref{eq:rad_power}), these fluctuations can be understood as Brownian motion of the trapped nanoparticle that is read-out optically via the intensity of the fiber's transmitted light.  The position of the nanoparticle can be extracted from the normalized power scattered by the nanoparticle using the relation $r_p=-k^{-1} \text{ln}[P_{\text{sca}}(r_p)]$~\cite{Arnold-2009}, where $k=2 \pi n_{\text{eff}}/ \lambda$ is the wavenumber and $n_{\text{eff}} = 1.3$ is the effective refractive index of a 500 nm HE$_{11}$ mode in water.  Following Ref.~\cite{Arnold-2009}, the maximum scattered power observed for a given trapping event is taken to conincide with the nanoparticle being in contact with the fiber.  Taking a histogram of $r_p$ then gives the probability distribution $p\left(r_p\right)$ of the nanoparticle's position, from which the potential energy of the trap $U(r_p)/k_B T = \text{ln} [p\left(r_p\right)]$ can be calculated from the equipartition theorem~\cite{Neuman-2004, Arnold-2009}.  The calculated trap potential is shown in Fig.~\ref{fig:nanospheres}(c) for a representative single nanoparticle event.  We numerically fit this data to a sum two exponentials (shown as the solid red line), which reveals an anharmonic trap with a depth of 5.5 $k_B T$, confining the nanoparticle approximately 78.4 nm away from the fiber surface.  We also calculate the expected gradient potential $U_{\text{grad}}(r_p)= -1/4 \epsilon_b \alpha |E(r_p)|^2$ (shown as the dashed black line) for our nanofiber, where the calculated field intensity is normalized for an input power of 810 $\mu$W.  Good agreement is found between the calculated potential and the numerical fit.

% August 7, 2013
% The figure in the paper with the potential data and fit is from the event between -60.92 s and -59.85 s.  The stiffness is 1.6 pN/um.  
% I took three random trapping events, and got stiffness of 1.6, 1.1 and 1.0.  So average of 1.2 +- 0.3 pN/um. 
% For these, I get max forces of 0.1 +- 0.008 pN and min forces of -0.5 +- 0.09 pN
% for the single event shown in the paper figure, its -0.6 pN and 0.1 pN

Since stable trapping is facilitated by a repulsive electrostatic force, the properties of the trap may be controlled by modifying the surface chemistry of the nanofiber.  We demonstrate this here by introducing 1 nM of NaCl into the nanoparticle solution.  Fig.~\ref{fig:nanospheres}(d) shows the intensity of transmitted light as a single nanoparticle diffuses into the optical field and is trapped near the fiber in NaCl solution.  The addition of NaCl reduces the electrostatic repulsion between surface groups~\cite{Arnold-2009}, thereby increasing the trap depth to 7.4 $k_B T$ and trapping the nanoparticle closer to the fiber surface (about 50 nm away).  In this case the duration of the optical trap was increased to 4.5 seconds and, in contrast to trapping without surface chemistry modification, the nanoparticle eventually binds to the fiber surface.  %Because the nanoparticle is trapped closer to the surface, the maximum value of the scattered power increased to about 833 nW, a factor of 5.6 greater than without NaCl.  Noting that the background noise was 36 nW over a 1 second sample, this corresponsd to a signal to noise ratio of about 23 for a single nanoparticle event.   

\begin{figure}[ht!]
%	\begin{center}
	\centering
	\includegraphics*[width=0.75\columnwidth]{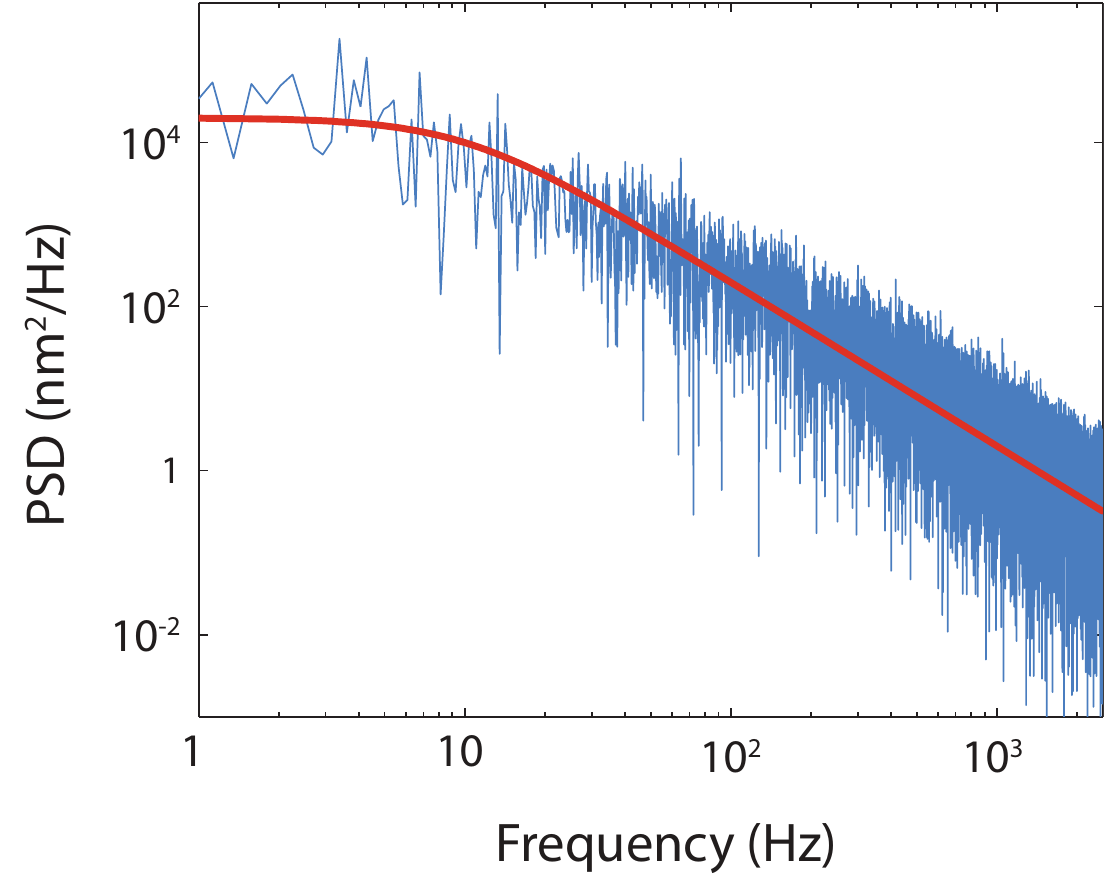}
	\caption{Power spectral density of position fluctuations (blue trace) for a single nanoparticle trapped in a solution of 1 nM NaCl.  The red curve is a Lorentzian fit which reveals a corner frequency of 10 Hz.}
	\label{fig:nanospheres2}
	%\end{center}
\end{figure}  

By differentiating the numerical fits to the trapping potentials with and without surface chemistry modification, we obtain the applied forces $F_{\text{trap}}(r_p)$ shown in Fig.~\ref{fig:nanospheres}(e).  From these we can obtain estimates of the trapping stability.  As expected, reduction of the electrostatic force via NaCl screening led to an increase in the optical gradient force, since the particle is trapped closer to the fiber surface.  We also observe an improvement in the trap stiffness, which is given by the slope of the force-distance curve at the position where the particle is trapped.  For purified water, the maximum gradient force was 0.10 pN and the stiffness was 1.3 pN $\mu$m$^{-1}$.  With NaCl, however, these values increased to 0.14 pN and 2.7 pN $\mu$m$^{-1}$, respectively.  It is interesting to compare these forces with those which are possible in a conventional optical tweezer trap (NA = 1) with identical particles and optical power.  Using the toolbox described in Ref.~\cite{Nieminen-2007}, we find that for these conditions the maximum trapping force is less than 0.01 pN.  Thus, the strong evanescent field of a nanofiber tweezer improves the optical trapping forces by a factor of approximately 14.  We note that the use of layered nanoparticles with a high-refractive-index core produces larger trapping forces than what is achievable with lower refractive-index probes such as polystyrene~\cite{Jannasch-2012}.  Moreover, the high-refractive-index of the nanoparticles used in this work results in a larger scattering cross-section through Eq.~\ref{eq:rad_power}, thus improving the optical read-out of the trapped particle's position.  The maximum electrostatic force applied in the nanofiber tweezer was -0.5 pN, a factor of 50 greater than that possible in conventional optical tweezers.   Since electrostatic forces scale with particle area,  while optical scattering forces scale with volume squared, one would expect electrostatic forces to scale much more favorably for smaller particle sizes.  This improved scaling could be relevant for biological situations in which both low optical power and strong trapping forces are desirable.  Furthermore, while we have carried out these experiments with Fe$_2$O$_3$ nanoparticles, even larger optical forces could be expected for plasmonic nanoparticles (e.g., gold) due to their enhanced polarizability~\cite{Swaim-2011}.  

%We also observe a slightly larger trapping stiffness of 2.7 pN $\mu$m$^{-1}$, which is comparable to the value achieved using a high-Q microresonator~\cite{Arnold-2009}, and 20$\times$ larger than what can be achieved with conventional optical tweezers for 500 nm polystyrene particles for our level of optical power~\cite{Erickson-2011}.  We believe this highlights the advantage of using nanofibers as simple, yet sensitive optical sensors for nano-scale particles.  

%  In this case, electrostatic screening from the addition of NaCl permits the nanoparticles to move closer to the surface and eventually bind due to attractive forces such as the induced dipole interaction.  By trapping the nanoparticle closer to the surface, the maximum scattered power from a single nanoparticle was increased to 833 nW, a factor of 5.6 greater than without NaCl.  Noting that the background noise was 36 nW over a 1 second sample, a signal to noise ratio of about 23 is achieved for a single nanoparticle event. 

Lastly, we calculate the power spectal density (PSD) of the position fluctuations from the nanoparticle trapped in Fig.~\ref{fig:nanospheres}(c).  Shown in Fig.~\ref{fig:nanospheres2}, the calculated PSD reveals a characteristic Brownian motion spectrum, with a corner frequency of about 10 Hz, which is well matched to a Lorentzian fit.  At 2.5 kHz, the position sensitivity is 3.4 $\AA / \sqrt{\text{Hz}}$, which is comparable to the best sensitivities previously reported for similar sized nanoparticles~\cite{Neuman-2008} and noteworthy given that it was achieved without optimization.  

% without the need for large changes in the injected optical power.  changing the opt, and control the position of the allow the electrosto match the electrostatic force and stably trap plasmonic nanoparticles.  Whereas the scattering and absorption forces are used as the restoring force in conventional optical tweezes, these forces are tangential to the driving field in nanofibers, and can be used to propel the nanoparticles along the direction of light propogation~\cite{Arnold-carousel}.

In conclusion, we have shown that the combination of attractive optical gradient forces and repulsive electrostatic forces allows trapping of nanoparticles with a tapered nanofiber.  Electrostatic forces are independent of optical power and scale with the nanoparticle's area, providing a better scaling of trap strength with nanoparticle size than that in conventional optical tweezers.  This allows trapping of particles with forces that are a factor of 50 larger than what is possible conventional optical tweezers.  We expect that nanofibers could find applications in single molecule force spectroscopy, where large forces are desirable to study complex biological structures and low power is needed to reduce the risk of damaging the sample.

The authors would like to acknowledge Jiangang Zhu for helpful discussions on tapered optical fibers, David Thompson for his knowledge of nanoparticle interactions, and the Australian Research Council for funding (Grant No. DP0987146).

\end{document}